\documentstyle[amssymb,aps,twocolumn]{revtex}

\begin{document}
\author{S. Ouazi$^{1}$ , J. Bobroff$^{1}$, H.\ Alloul$^{1}$, W.A. MacFarlane$^{1}$}
\title{Correlation length in cuprate superconductors deduced from impurity-induced
magnetization}
\address{$^{1}$LPS, UMR 8502, 91405\ Orsay Cedex, France}
\maketitle

\begin{abstract}
We report a new multinuclei based{\bf \ }nuclear magnetic resonance method
which allows us to image the staggered polarization induced by nonmagnetic
Li impurities in underdoped (O$_{6.6}$) and slightly overdoped (O$_{7}$) YBa$%
_{2}$Cu$_{3}$O$_{6+y}$ above T$_{C}$. The spatial{\bf \ }extension of the
polarization $\xi _{imp}$ approximately follows a Curie law, increasing up
to six lattice constants at $80K$ at O$_{6.6}$ in the pseudogap regime. Near
optimal doping, the staggered magnetization has the same shape, with $\xi
_{imp}$ reduced by a factor $2$. $\xi _{imp}$ is argued to reveal the
intrinsic magnetic correlation length of the pure system. It is found to
display a smooth evolution through the pseudogap regime.
\end{abstract}


\bigskip I- INTRODUCTION

\bigskip

In high Tc cuprates, strong antiferromagnetic (AF) correlations reminiscent
of the AF\ ordered phase were often argued to explain the anomalous normal
state properties of the CuO$_{2}$ planes and to lead to the d-wave pairing
in the superconducting state \cite{Kampf}. Nuclear magnetic resonance (NMR)
and neutron scattering experiments both showed the existence of such
correlations and their decrease with increasing doping. However, no
agreement has been reached about quantitative determinations of the AF\
correlation length $\xi $ values\ and about a possible incommensurability.
In underdoped YBa$_{2}$Cu$_{3}$O$_{6+y}$, some neutron experiments detect a
magnetic response peak with a temperature ($T$) independent $q$ width, which
corresponds to $\xi \approx 2.5\ $(in lattice units) \cite{BourgesBalatski}%
.\ But this response could be dominated by a macroscopic inhomogeneity in
the large single crystals used or according to other experiments an
incommensurability, so that $\xi \;$could be underestimated.\ In contrast,
the analysis of Pines {\it et al. }(noted hereafter MMP) of the Cu NMR
relaxation times T$_{1}$ and T$_{2}$ suggests a Curie-Weiss behavior for
\thinspace $\xi \;$which might reach values as large as $7$ at $100K$ \cite
{BarzykinPines}. But, various reexaminations of the MMP analysis lead to
substantial uncertainties in their $\xi $ values, so that at O$_{6.6}$, a
model-free determination of $\xi $ versus $T$ is still lacking. Near optimal
doping, the magnetic response peak measured by neutrons broadens a lot and
its intensity is reduced indicating smaller magnetic correlations. This
makes the experimental data difficult to analyse quantitatively.

We propose to use nonmagnetic impurities to determine this length scale $\xi 
$ in cuprates. Nonmagnetic substitutions on the magnetic sites of low
dimensionnal AF correlated systems are known to lead to the appearance of an
AF staggered magnetization. This has been observed by NMR in cuprates \cite
{Mahajan}\cite{JBobroff1997}\cite{MarcHenriJ}, spin chains \cite{Tedoldi},
and spin ladders \cite{TakigawaFujiwara}. In the case of an AF $S=1$ spin
chain, the induced polarization could be imaged site by site \cite{Tedoldi}%
.\ Its extension was found to be identical to the correlation length of the
pure system. In the same spirit, we propose a detailed quantitative study of
the staggered magnetization induced by nonmagnetic Li and Zn substituted at
the Cu sites of CuO$_{2}$ planes in YBa$_{2}$Cu$_{3}$O$_{6+y}$. Previous
studies proved that the staggered polarization has a non-Fermi liquid $T$
dependence, confirming the correlated nature of the planes \cite
{JBobroff1997}. However, no quantitative determination of the extent or
shape of this polarization could be achieved.{\em \ }It is not yet clear
whether the decay of the spin polarization can be described by a single
length scale.\ One can indeeed wonder whether charge and spin degrees of
freedom can result in different responses. Here, we measure the
modifications induced by nonmagnetic impurities on the NMR spectra of all
the available nuclei, $^{17}$O, $^{89}$Y and the $^{7}$Li impurity itself.\
Combining all these informations we determine, for the first time, the shape
of the impurity induced polarization on a significantly large lengthscale
and deduce that a single lengthscale $\xi _{imp}\;$can be used.

This ''multinuclei'' analysis allows us to deduce the spatial variation of
the polarization with temperature and doping. We demonstrate that $\xi
_{imp} $ increases with decreasing temperature in the pseudogap phase.\ This
polarization persists, but is reduced in O$_{7}$, out of the pseudogap
phase, together with the Kondo-type $T$ variation of its magnitude evidenced
in \cite{JBPRLLi}. Comparison with the MMP analysis suggests that $\xi
_{imp} $ is, as in spin chains, a direct measure of the correlation length $%
\xi \;$of the pure system.\ 

The paper is organized as follows: Sec. II is devoted to the description of
samples and of the NMR experimental conditions. In Sec. III we decribe the
principle of our multinuclei method. In Sec. IV, we present the experimental
results. The next two sections are concerned with the description and
analysis of the experimental results. We detail in Sec. V the steps for
determining the induced magnetization and present our results for the
induced magnetism at O$_{6.6}$ and O$_{7}$. A comparison with neutron
experiments in Zn substituted O$_{7}$ is discussed in Sec. VI. We compare
also quantitatively our findings with the neutron scattering and nuclear
spin relaxation measurements in the pure YBCO compound.

\bigskip

II- EXPERIMENTAL DETAILS

\bigskip

NMR measurements presented hereafter were done on YBa$_{2}($Cu$_{1-x_{N}}$Li$%
_{x_{N}}$)$_{3}$O$_{6+y}$ compounds with $x_{N}=0$, $0.01$ and $0.02$.
According to measurements done in \cite{JBPRLLi} on the same samples, a
quantity $x_{plane}=0.85\times x_{N}$ of Li is found to substitute the Cu
site of the planes. For all the samples, the $^{17}$O enrichment procedure
was carried out as in \cite{JBobroff1997}. The samples were fully reoxidized
at $300%
{{}^\circ}%
C$ leading to the slightly overdoped composition $y=1$ (noted hereafter as O$%
_{7}$).\ Deoxidation of part of each batch allowed to produce the underdoped
samples $y=0.6\;$(O$_{6.6}$) under thermogravimetric control. The powders
were finally aligned in Stycast epoxy in a high magnetic field so that their 
$c\;$axis could then be chosen to lie parallel to the NMR static field.\ Our
samples present a misorientation of less than 10$%
{{}^\circ}%
\;$of$\;$the c axis which introduces a second order quadrupolar broadening
of the Cu spectra but has negligible effect on the $^{89}$Y spectrum and on
the\ central transition of the $^{17}$O NMR. We performed $^{17}$O, $^{89}$%
Y, $^{7}$Li measurements at O$_{7}$ and O$_{6.6}$ (some of the $^{89}$Y\ and 
$^{7}$Li NMR data at O$_{6.6}$ were already published in \cite{JBPRLLi}).
Spectra were obtained using Fourier Transform of the echoes obtained with a
standard $(\pi /2)-\tau -\pi $ sequence under typical fields $H_{ext}$ of $7$
Tesla. We also performed $^{63}$Cu NMR measurements in the underdoped
compound at $300K$ and $100K$ in a $5$Tesla applied field.

\bigskip

III- MULTINUCLEI METHOD

\bigskip

The key point of our multinuclei method is that each nuclear species is
coupled differently to the electronic spin density on the Cu sites, i.e.,
they have different hyperfine form factors \cite{MilaRice}. This was already
judiciously used in the pure system to distinguish between uniform and
antiferromagnetic spin fluctuations through T$_{1}$ measurements \cite
{BarzykinPines}. We proceed in the same spirit for the impurity problem.\ We
assume that an impurity positioned at the origin induces a polarization $%
\langle S^{z}(\overrightarrow{r},T)\rangle $ noted $S_{n}^{m}$ on the Cu
site at $\overrightarrow{r}=(m,n)$ in lattice units from the impurity. The $%
^{17}$O nucleus is predominantly coupled to the spin density of its two
copper neighbors and has an NMR resonance shifted by :

\[
^{17}K(m,n)=^{\,\;17}\alpha (S_{n}^{m}+S_{n}^{m+1}) 
\]

where $^{17}\alpha =^{17}A_{hf}g/H_{ext}$, $^{17}A_{hf}$ is the hyperfine
coupling between $^{17}$O and one Cu site in the $c$ direction of the
measurements and $g$ is the Land\'{e} factor. The other nuclei follow in the
same manner: 
\begin{eqnarray*}
^{89}K(m,n) &=&^{89}\alpha (S_{n}^{m}+S_{n}^{m+1}+S_{n+1}^{m}+S_{n+1}^{m+1}),
\\
^{63}K(m,n) &=&^{63}\alpha (\frac{A}{B}%
S_{n}^{m}+S_{n}^{m+1}+S_{n}^{m-1}+S_{n+1}^{m}+S_{n-1}^{m}), \\
^{7}K(0,0) &=&^{7}\alpha \ast 4\ast S_{0}^{1}
\end{eqnarray*}
where $A$\ and\ $B$\ are respectively the Cu on-site and nearest neighbor
supertransferred hyperfine couplings \cite{hyperfin}. Due to its specific
position, the $^{7}$Li shift probes only the polarization of the $r=1$\ Cu
near neighbors (noted nn).\ For the other nuclei, the different geometrical
factors act like filters. To illustrate this effect, in Fig.1 we plot the
magnitude of these shifts $\mid K(m,0)\mid $ along the $(1,0)$ direction and
on checkerboards for an exponentially decaying polarization. \bigskip For an
alternated polarization, the $^{17}$O shift $^{17}K(m,n)\;$measures a
quantity which roughly corresponds to the $1st$ derivative of the
polarization envelope. Similarly, the $^{89}$Y shift $^{89}K(m,n)$ rather
represents the curvature of the polarization envelope. The more complex
filtering operated by the shift measurement of $^{63}$Cu combines both a
direct contribution of the spin polarization and of its derivatives. From
these couplings, it can be intuitively understood that the various nuclei
will not be sensitive to the same range of the polarization. This can be
seen in Fig.\ 1, where, for a given polarization, we represent the local
shifts using the proper filter for each nucleus. Furthermore, for each
nuclear species, the existence of various sources of NMR line broadening in
the pure materials limits the experimental sensitivity and defines a cut-off
on the distance which can be probed. This leads us to conclude that $^{89}$Y
primarily probes the nearfield ($r<3$) of the polarization whereas $^{17}$O
is sensitive to $r\lesssim 8.\;$The comparison between NMR spectra of these
different nuclei should then allow\ to recover an information on a large
range of distances. We note that $^{63}$Cu NMR in principle probes the full
range and might thus appear the ideal probe to determine the whole
polarization.\ Unfortunately the experimental conditions are not sufficient
to exploit the $^{63}$Cu\ NMR spectra for a quantitative analysis, as will
be detailed hereafter.

\bigskip

IV- \bigskip EXPERIMENTAL RESULTS

Our measurements show indeed qualitatively different features for these
various nuclei. This confirms that these nuclei do not probe the same
spatial ranges of the Cu spin polarization. In underdoped O$_{6.6}$ samples,
two well separated satellites appear in the $^{89}$Y NMR spectra as shown in
Fig. 2. They correspond to the first and second nn to the Li$\ $as
anticipated from Fig. 1. The $^{17}$O NMR spectra do not show such
satellites, but are symetrically broadened by the impurities as shown in the
inset of Fig. 3. This broadening has a faster $T-$variation than that of the 
$^{89}$Y satellites shift which follows a Curie-like variation.{\em \ }The $%
^{7}$Li NMR\ shift also follows a $1/T$ Curie law in the $T-$range measured
here \cite{JBPRLLi}. As for the $^{63}$Cu NMR , the large second order
quadrupole effects makes it extremely sensitive in pure samples on the
preparation conditions and on the quality of the alignement of the powder.\
So it is hard to compare accurately the spectra of pure and substituted
samples.We have however measured between $300K$\ and $100K$\ an absolute
increase of the linewidth of $23000\pm 4000$ppm for $2\%$ Li.\ Although
these numbers will be useful for comparison with the data on the other
nuclei, we considered that a systematic study of the Cu NMR in our specific
experimental conditions would not be useful.\ This will be discussed further
in Sec. V B 1.

In slightly overdoped O$_{7}$, the impurity effects on the NMR spectra are
not only strongly reduced but also display a qualitative change in their$\;T$
dependences. Discrete NMR satellite resonances are no more detectable in the 
$^{89}$Y spectra which only exhibit a slight broadening which varies roughly
as $1/T$ (see the inset of Fig. 2).{\bf \ }The $^{17}$O width also increases
as $1/T$ (see Fig. 3) while the $^{7}$Li shift follows a $1/(T+135)$
Curie-Weiss behavior. The behaviors of the $^{89}$Y and the $^{17}$O spectra
both at O$_{6.6}$ and O$_{7}$ are qualitatively very similar to those
observed in Zn-substituted compounds. Their magnitude only differ by at most 
$20\%$ from that observed on Li substituted samples for the same in-plane
impurity concentration. Therefore, the observations done for Li impurities
can be considered as universal signatures of nonmagnetic impurity effects in
cuprates.

\bigskip

V- DETERMINATION OF THE INDUCED MAGNETIZATION

\bigskip

We now apply the multinuclei analysis to these findings in order to
reconstruct the induced polarization. We performed extensive numerical
simulations in order to fit the data, constrain the shape of the
corresponding polarization, and deduce its amplitude and the $T$ dependence
of its extension. Our methodology is detailed in Sec. V A, in the regime
where the measurements show the largest impurity-induced changes, i.e., in
the O$_{6.6}$ compound at the lowest temperatures. We demonstrate that a
single length scale and shape actually describes the spatial dependence of
the polarization.\ In Sec. V\ B this methodology has been systematically
applied versus temperature.\ It allows us to present our main results, i.e.,
the spatial characterization of the induced magnetism from an underdoped
compound O$_{6.6}$ up to a slightly overdoped compound O$_{7}$.\ 

\bigskip

{\bf A. Methodology }

\bigskip

To get quantitative information from these experimental observations, let us
write the local AF\ polarization as such:

\begin{equation}
\langle S^{z}(\overrightarrow{r},T)\rangle =-S_{0}^{1}\ (-1)^{m+n}f_{T}(r)
\end{equation}

where $f_{T}(\overrightarrow{r})$ is the envelope of the staggered
polarization at a temperature $T$ which is normalized to $f_{T}(%
\overrightarrow{r})=1$ for $\parallel \overrightarrow{r}\parallel =1$.

The hyperfine couplings of the pure compounds are known for all nuclei \cite
{hyperfin} and can be retained for nuclear sites which are not nearest
neighbors of the impurities. However, the hyperfine coupling $%
^{89}A_{hf}^{1}\;$between the $^{63}$Cu and $^{89}$Y nearest neighbors of
the same Li impurity is expected to be modified.\ The $^{7}$Li hyperfine
coupling $^{7}A_{hf}$ is not knwown accurately as well. The analysis done
below will allow us to constrain their values.

We have performed numerical simulations of the $^{89}$Y and $^{17}$O spectra
for various choices of amplitude $S_{0}^{1}$ and shape $f_{T}(%
\overrightarrow{r})$ on a checkerboard\ of $150\ast 150$ cells, with a
concentration $x_{plane}$ of randomly distributed impurities.

In Sec. V A 1, we demonstrate that the experimental features of the $^{17}$O
and $^{89}$Y \ NMR can only be accounted for by a continuous exponentially
shaped polarization and hence by only one length scale. In Sec. V A 2 the $%
^{17}$O spectrum shape further allows us to determine $S_{0}^{1},\;$and
therefore $^{89}A_{hf}^{1}$ and $^{7}A_{hf}.\;$

\bigskip

\ \ \ \ \ \ \ \ {\it \ 1. Shape and continuity of the spin polarization}

\bigskip

At a given temperature $S_{0}^{1}\;$is initially unknown and the $^{89}$Y
first satellite NMR\ shift is not useful here to analyse the polarization
shape.\ At O$_{6.6}$, the final amplitude $S_{0}^{1}$\ and shape $f_{T}(%
\overrightarrow{r})$ must account for (i{\it )} the presence of two $^{89}$Y
satellites on the spectrum, (ii{\it )} the shift of the second $^{89}$Y
satellite and (iii{\it )} the Lorentzian like shape of the $^{17}$O\ line.

We shall illustrate on the $80K$\ data the analysis done at various
temperatures. First we use the experimental features of the $^{89}$Y
spectrum to constrain the polarization envelope at short distances from the
impurity. The assumption of a given amplitude $S_{0}^{1}$ at $80K$\ leads to
a unique choice for $f_{T}(\overrightarrow{r})$\ for which only two $^{89}$Y
satellites distinct from the central line exist. This can be clearly seen in
Fig. 4. For a given amplitude $S_{0}^{1}$, different curvatures $A$, $B$ and 
$C$ lead to $^{89}$Y spectra with qualitatively different spectral
structures. Only one of them ($B$) presents two satellites and fit the
second satellite shift. The decay $C$ is too smooth to produce two
satellites distinct from the central line. The sharper curvature $A$ leads
to a third satellite not observed experimentally. Moreover, this
polarization shape cannot account quantitatively for the second satellite
position. On the contrary, the exponentially decaying $B$\ accounts well for
both the presence of two satellites and the value of the shift of the second
satellite. Hence $^{89}$Y data allows to constrain $f_{T}(\overrightarrow{r}%
) $ up to the fourth Cu sites ($r\leq \sqrt{5}$) for a given choice of $%
S_{0}^{1}$. Anyhow, different solutions for different $S_{0}^{1}$ are still
compatible with the experiments. This is exemplified in Fig. 4 where both $B$
and $D$ polarizations represented in the inset of Fig. 4 lead to $^{89}$Y
spectra with only two satellites and the second one at the proper position.

In order to constrain the shape $f_{T}(\overrightarrow{r})$ at further
distances, we use now the $^{17}$O spectra. Let us consider as an example
the short range polarization envelope represented by the curve $B$ in Fig.
4. For $r\geq \sqrt{5}$, we free the shape in order to account for the
experimental spectrum. Figure 5 represents simulated $^{17}$O spectra and
their corresponding polarizations. The only satisfying fit is obtained for
an exponential curvature which prolongates the short distance decay $B$. A
polarization with two different length scales results in a more complex
spectral shape. This is exemplified by the simulated spectra in Fig. 5. The
experimental smooth lorentzian spectrum can only be explained by a
continuous exponentially shaped envelope. On the spectra corresponding to
discontinuous polarizations, a satellite or shoulders appear on the $^{17}$O
spectra . This allows then to exclude the scenario in which a large induced
moment appears on a short length scale and yields a longer range spin
polarization.

\bigskip \bigskip

\ \ \ \ \ \ \ \ \ \-{\it 2. Amplitude}

\bigskip {\bf \ }

To determine now the polarization amplitude $S_{0}^{1}$, we attempt fits of
the experimental $^{17}$O spectrum with the various possible short range
polarizations which correspond to distinct $S_{0}^{1}$ such as $D$ and $E$
(see Fig. 6). We let the polarization shape free for $r\geq \sqrt{5}$. The
Lorentzian $^{17}$O experimental spectrum can be fit at best by the
represented curvatures. The shape required for $r\geq \sqrt{5}$ is again
found exponential and continuous with the short range curvature. But all
these cases do not fit the $^{17}$O spectrum equally well. Case $D$ results
in an $^{17}$O spectrum with shoulders and case $E$ can not explain
quantitatively the broadening. Eventhough $D$, $B$ and $E$ all account for $%
^{89}$Y spectra only B accounts as well for the $^{17}$O {\it shape and width%
} at the same temperature.

Thus $^{17}$O NMR allows to select the proper amplitude $S_{0}^{1}$, and
therefore yields a value for the Li hyperfine coupling $^{7}A_{hf}$ through $%
S_{0}^{1}(T=80)=^{7}K(T=80K)$ $H_{ext}/4^{7}A_{hf}g$. Taking into account
the experimental accuracy, we found that $^{7}A_{hf}$ ranges from $0.85$ to $%
1.05kOe$. This result is consistent with our previous upper experimental
estimation $^{7}A_{hf}<$ $2.5kOe$ \cite{JBPRLLi}, done from the comparison
with the measured susceptibility. The value $S_{0}^{1}$ also allows us to
determine the $^{89}$Y first satellite shift and the first nn\ $^{89}$Y \
hyperfine field $^{89}A_{hf}^{1}=0.8\times ^{89}A_{hf}$, which only slightly
differs from that of the bulk.

For increasing $T$ , the $^{89}$Y experimental spectrum always presents two
satellites and the $^{17}$O lineshape remains Lorentzian. Hence the
polarization envelope shape has to be exponential at all temperatures, and a
single length scale $\xi _{imp}$ is necessary to describe it. In the
following, we choose the Bessel functional form $f_{T}(r)=K_{0}(r/\xi
_{imp})/K_{0}(1/\xi _{imp})$,\ which approximates an exponential $%
exp(-2r/\xi _{imp})$ for the experimental range of values of $r/\xi _{imp}$\
explored. This mathematical notation is used here rather than the
exponential one as it is exactly that which appears in the MMP analysis of
the relaxation data. The fact that the same analytical definition is taken
for $\xi _{imp}$ and $\xi $ in the two approaches will allow us to compare
them quantitatively.

\bigskip

{\bf B. Results}

\bigskip \bigskip

As shown above the experimental data constrains a Bessel-type polarization
shape at each temperature. Its extension $\xi _{imp}$ is then directly
determined by the $^{17}$O impurity-induced broadening.

\bigskip

\ \ \ \ \ \ \ \ \ {\it 1. Induced magnetism in underdoped O}$_{{\bf 6.6}}$

\bigskip

The corresponding findings for $\xi _{imp}(T)$ are presented in Fig. 7, the
upper (lower) bounds of the error bars being associated with the higher
(lower) limit for $^{7}A_{hf}$.\ Our central results consists then in the
restricted range of a couple of values for the amplitude $S_{0}^{1}$ and the
extension $\xi _{imp}(T)$ of the staggered polarization.\ From the obtained
values of $\xi _{imp}(T)$,$\;$we could compute the $^{89}$Y satellites
positions over the full range of temperatures.\ Those are{\em \ }found to
agree reasonably well with the experimental ones as can be seen in Fig. 2.
This confirms {\it a posteriori} the validity of our analysis and of the
value obtained for $^{89}A_{hf}^{1}$ .

To check further the coherence of the analysis we computed the corresponding 
$^{63}$Cu spectra and compared them to the experimental ones. At $100K$, our
simulations lead to a $^{63}$Cu broadening between $22000$ppm and $25000$%
ppm. This value is very similar to our measured broadening. But a
quantitative comparison between the experimental Cu spectra and the
simulated ones requires to pay further attention to some experimental
complications.{\it \ }Recent Nuclear Quadrupole Resonance\ experiments
evidenced that nuclei up to at least the fourth nn of the impurities are
undetectable due to relaxation effects or to an impurity contribution to the
electric field gradient \cite{ITOH}. As such effects should then occur as
well in our NMR measurements, we measured the $x_{N}$ dependence of the $%
^{63}$Cu NMR intensity{\em .} Assuming that a nucleus either fully
participates to the NMR line, or is merely wiped out, we estimated a cutoff
radius of about four unit cells (slightly smaller than that of \cite{ITOH}).
We then computed the expected impurity-induced broadening of the Cu spectrum
when this wipe out is taken into account. This broadening is reduced to a
value between $16000$ and $18000$ppm, slightly smaller than the experimental
value of $23000$ppm. Other $^{63}$Cu measurements have been done on a single
crystal sample with a similar Zn concentration but with slightly larger hole
doping \cite{MarcHenriJ}. They show an impurity-induced broadening of $12600$%
ppm at $100K$, almost $2$ times smaller than ours.{\em \ }For this data
taken in different experimental conditions (external field, sample doping,
and dopant), comparisons beyond order of magnitude agreement cannot be
performed.

To conclude this comparison with $^{63}$Cu NMR data an overall agreement is
obtained, but the experimental complications mentioned above prevents
getting any further information from the NMR spectra in our experimental
conditions.

\bigskip

\ \ \ \ \ \ \ \ \ {\it 2. Induced magnetism in slightly overdoped O}$_{{\bf 7%
}}$

\bigskip

As the hyperfine couplings are usually found independent of doping in
cuprates, the value of $^{7}A_{hf}\;$determined from the O$_{6.6}$\ analysis
can be kept, so the data for $^{7}K(T)$ directly yield the values of $%
S_{0}^{1}$. Hereagain, the Lorentzian shape of the $^{17}$O NMR\ line
constrains an exponential shape for the polarization envelope, and the
impurity-induced broadening directly allowed us to deduce $\xi _{imp}(T)$,
which is plotted in Fig. 7. The corresponding $^{89}$Y simulated satellite
positions are plotted in the inset of Fig. 2. They appear much less shifted
from the main line than at O$_{6.6}$.\ This is expected, as the Li shift
indicates that the polarization on the first nn Cu is much smaller than in O$%
_{6.6}$. A convolution with the typical lineshape of the pure sample allows
us to confirm that such satellites cannot be resolved in agreement with the
data, but contribute to the broadening of the $^{89}$Y line which fits the
experimental one.

\bigskip

\ \ \ \ \ \ \ \ \ {\it 3. Comparison with the macroscopic susceptibility data%
}

\bigskip

Finally we compute the impurity contribution to the macroscopic
susceptibility $\chi _{imp}$ which is given by the sum of $\langle S^{z}(%
\overrightarrow{r})\rangle $ on all Cu sites of the checkerboard. At O$%
_{6.6} $, the computed $\chi _{imp}$ follows a Curie-Weiss law in $%
p_{eff}^{2}/(T+\Theta )$ with $\Theta <100K$ and an effective magnetic
moment $p_{eff}=1.5(3)\mu _{B}$ per impurity. This behavior agrees within
our uncertainties with the Curie law measured in Zn-substituted high purity
compounds \cite{PMendels}. At O$_{7}$, the calculated $\chi _{imp}$\ is also
compatible with these data, as it presents a strong reduction mostly due to
an increase of $\Theta $. This confirms the robustness of our analysis. This
variation of the Curie -Weiss temperature detected on the first nn has been
attributed to a Kondo-type screening of the induced moment by the carriers 
\cite{JBPRLLi}. From Eq. (1), we find here that this Kondo-type $T$
dependence merely multiplies the whole AF response. This situation is
similar to the standard Kondo effect in dilute alloys where $\langle S^{z}(%
\overrightarrow{r},T)\rangle =\langle S_{impurity}^{z}(T)\rangle f(r)$ \cite
{BoyceSlichter}. However in our case the impurity itself is not magnetic and
the function $f(r)$\ actually depends on $T$\ as a consequence of the
variation of $\xi $.

\bigskip \bigskip

VI- DISCUSSION

\bigskip

\bigskip {\bf A. Comparison to other impurity studies }

\bigskip

Let us compare now our findings to the existing experimental studies. Our
induced magnetization model is compatible with previous NMR measurements 
\cite{Mahajan} \cite{JBobroff1997} \cite{MarcHenriJ} in YBCO with
nonmagnetic Zn impurities. Inelastic neutrons scattering (INS) measurements
performed on O$_{7}$ + Zn have evidenced that Zn substitutions enhance low
energy AF fluctuations \cite{Sidis}. The $q\ $width of the inelastic peak at
the AF wave vector $Q_{AF}$ reveals a length scale $\xi _{n}$ of $1.3$ cell
units at $150K$ and $1.6$ cell units just above T$_{c}$. This length scale
obtained either with a Lorentzian or Gaussian fit of the peak is very
similar to our result (Fig. 8). The neutron peak results from the
superposition of the impurity induced AF fluctuations on the contribution of
the pure system. It is then difficult to attribute a simple significance to $%
\xi _{n}$. On the contrary, in our analysis, we deal only with the
impurity-induced magnetism. The quantitative agreement between these results
is therefore quite surprising but might indicate that these length scales
are not that different, at least for O$_{7}$. On the other side, the
longitudinal spin lattice relaxation $T_{1}$ of $^{7}$Li \cite{PRLAndrew}
relates directly to the low frequency response probed by INS.\ The present
determination of $^{7}A_{hf}$ \ allows us to estimate the electronic
fluctuation time $\tau \;$of the nn induced magnetic moment.\ It corresponds
to an energy scale\ $\ h/\tau =$ $(18.4\pm 3.7)meV$ at $100K$ in O$_{7}$
(independently of Li\ concentration) \cite{footnoteTAU} very similar to the
value $15\pm 3meV$ observed by INS in O$_{7}$ with $1.6\%$ Zn\ \cite
{theseSidis}.\ Thus at O$_{7},$ in presence of $1.6\%$ Zn, the fluctuation
time of the induced moments is also coherent with the energy of the
impurity-induced AF fluctuations measured by INS. We however stress here
that the present NMR experiments allow to separate unambiguously the
impurity-induced magnetism from the pure one. Our method gives the spatial
dependence of the staggered magnetization and its temperature dependence,
and allowed us to probe as well the induced magnetism in the underdoped
regime.

Various theoretical studies show that the AF correlations in the pure
compound are responsible for the appearance of an AF polarization near a
spin vacancy. Weak coupling AF spin fluctuation models lead to a
polarization similar to ours within RPA \cite{WeakCorrelations}. Strong
coupling models such as t-J or Resonating Valence Bond also anticipate the
observed polarization, at least qualitatively \cite{StrongCorr}. However,
contrary to the case of undoped spin chains none of these approaches have
attempted to establish a relation between the extension of the induced
magnetism and the correlation length in the pure compound.

Let us now compare the polarization extension $\xi _{imp}\,$near the
impurities to the correlation length $\xi $\ as obtained from experimental
studies on the pure compounds.

\bigskip

{\bf B. Comparison to the MMP analysis }

\bigskip

Indirect information can be obtained through the analysis of T$_{1}$ and T$%
_{2}$ \ $^{63}$Cu NMR measurements as done by MMP. At O$_{7}$, the MMP
analysis yields a nearly constant $\xi \simeq 2$ between $100K$ and $200K$,
represented on figure 8. Within the error bars, this value is very close to
our value of $\xi _{imp}$.\ In contrast, in the underdoped regime, we find a
value of $\xi _{imp}\,\;$at room $T$ about 5 times smaller than that deduced
by MMP$\;$for $\xi $, and a much larger $T$\ dependence (see figure 8).\
However, the MMP analysis relies on $T_{2}$ which can be strongly affected
by the occurence of the intrabilayer coupling and also by a possible
incommensurability. These effects both lead to an overestimate of $\xi $,
especially at O$_{6.6}$\cite{MorrIncomT2} \cite{EffetplansT2}. Hence, the
only reliable estimates from the MMP analysis are those done for O$_{7}$.\
At this doping, the similarity between $\xi _{MMP}$ and our $\xi _{imp}\;$is
therefore a good indication that the impurity reveals the correlation length
of the pure compound.\ The much better experimental accuracy obtained for $%
\xi _{imp}(T)$ as compared to $\xi _{MMP}$ implies that $\xi $ nearly varies
as $1/T$ at O$_{7}$. At O$_{6.6}$, our determination of $\xi _{imp}(T)$\
should then presumably also represent the $T$ variation of $\xi $.\bigskip

\bigskip

{\bf C. Comparison to INS measurements in the pure compounds}

\bigskip

INS experiments in pure compounds in principle allow to probe $\xi $ through
the $q\ $width of the AF peak ($(1/\Delta q)\varpropto \xi $), when
detectable. Incommensurabilities of the AF\ response might yield a
broadening which leads then to an overestimate of $\Delta q$. Hence neutrons
usually give a lower bound for $\xi $. Our results are compatible with
previous INS results if taking into account a possible incommensurability or
energy dependance of the neutrons data (see Fig. 8, in \cite{BourgesBalatski}%
). But the absence of any $T$\ variation of the AF peak seen by INS {\it a
priori} contrasts with our observations.

First, these INS experiments probe high energy fluctuations, contrary to NMR
which measures the zero-energy limit. Aeppli {\it et al.} in pure LSCO
(almost optimally doped \cite{AeppliLSCO}) showed the existence of an energy
dependence for $\Delta q$. When the energy is lowered, $\Delta q$ is reduced
at low $T$ and enhanced at high $T$. This suggests that the correlation
length measured at low energy could still behave similarly to our data for $%
\xi _{imp}.${\em \ }A recent theoretical work \cite{Markiewicz} proposes a
spin excitation function $\chi "(q,\omega )$ with a small plateau around $%
Q_{AF}$ which would reconcile as well the $T$-independent $q\ $width of the
neutrons peak in the pure system with our $T$-dependent $\xi _{imp}$.

Of course an incommensurability $\delta $ of the magnetic response, as
evidenced in pure underdoped compounds by some neutrons experiments \cite
{Mook} might as well explain this apparent discrepancy.\ We considered
whether such a value for $\delta \,$ would play a role in the analysis of
our results.\ This should yield a large-range modulation of $f_{T}(r)$\
which should vanish at a distance of about eight \ or nine unit cells.
Fortunately,{\em \ }this effect does not affect the staggered magnetism at
short distances probed by $^{89}$Y and $^{17}$O.\ We found that the changes
induced in our simulated $^{17}$O\ spectra would not be detected within
experimental accuracy. So our determination of $\langle S^{z}(r,T)\rangle $\
for $r\lesssim 8$\ is therefore unaffected. In contrast $^{63}$Cu nuclei
probe the polarization at larger distances than $^{89}$Y and $^{17}$O and
might be sensitive to such an incommensurability.\ But probing the staggered
magnetization at $r>8$\ might only be ensured after securing that the
observations are not affected by interactions between impurities.
Determining the shape of the tail of the polarization with $^{63}$Cu NMR
thus requires a much better experimental resolution, i.e., high quality
single cristals with $x_{plane}$\ $<<1\%$.

In summary, we have compared the impurity induced polarization extension $%
\xi _{imp}(T)\;$to the correlation length $\xi (T)\;$estimated in the pure
compounds.\ We found that these quantities are similar to that obtained from
the MMP analysis in the optimally doped samples.\ Various possible sources
for a broadening of the INS AF\ peak in the underdoped samples have been
proposed.$\;$This lets us suggest that the spatial range $\xi _{imp}(T)$ of
the induced magnetization around an impurity might be the best estimate of $%
\xi (T)$.

\bigskip

VII-CONCLUSION

\bigskip {\it \ }{\bf \ \ }

In conclusion, we have used NMR to determine the shape and magnitude of the
staggered polarization induced by a nonmagnetic impurity.\ We have found
that the decay of the polarization is characterized by a single length
scale, and is best fitted by an exponential.\ This detailed information
obtained gives a new basic constraint on any microscopic model of cuprates.
This experiment allows us to demonstrate that the AF correlations in the
normal state of YBCO persist up to optimal doping with no qualitative abrupt
change when crossing the pseudogap line. The extension $\xi _{imp}$ of the
staggered polarization is found sizeable and $T$ dependent in both the
underdoped and slightly overdoped regimes.\ We suggest that this quantity
might be the best estimate of the actual correlation length in the pure
systems, although further theoretical support for this proposal is still
required.\ 

\bigskip

\bigskip ACKNOWLEDGEMENTS

\bigskip

The authors would like to thank G.\ Aeppli, P.Bourges, B.\ Keimer, P.
Mendels and Y.\ Sidis for constructive discussions.

\bigskip 

FIGURE 1 : Magnitude of the local shifts $\mid K(m,n)\mid $ at the $^{17}$O
(b), $^{89}$Y (c) and $^{63}$Cu (d) nuclei for $H//c$ from an
impurity-induced polarization $S_{n}^{m}$ as represented in absolute values
in (a) as function of the distance to the impurity along the $(1,0)$
direction. The checkerboards (insets) show the same quantities on the
bidimensionnal CuO$_{2}$ layer. Black color is for the maximum intensity and
each square corresponds to one nucleus. In (b), the $^{17}$O neighbor sites
to the impurity (hatched area) - not observed up to now - are not considered
here as they only sense one Cu site, and should undergo a large shift.

FIGURE 2 : Data for the shift with respect to the main $^{89}$Y line of the
two satellites resonances detected at O$_{6.6}$. A reference spectrum taken
at $100K$ is displayed. The computed satellite shifts with the parameters
values given in Fig. 6 are shown as gray lines. The inset displays the
equivalent figure for the O$_{7}$ case, and shows that the satellites cannot
be resolved. 

FIGURE 3 : Broadening of the NMR spectrum of the planar oxygens induced by
Li impurities with $x_{N}=2\%$ for O$_{6.6}$ (triangles) and O$_{7}$
(circles). The broadening of the line in pure compound has been substracted.
In the inset is displayed a typical Lorentzian spectrum for O$_{6.6}$ with $%
x_{N}=2\%$ Li at $80K$. The small ''bump'' in the low frequency tail
corresponds to the resonance line of the apical site, almost completely
attenuated by dynamical contrast (fast repetition time). 

FIGURE 4 : Comparison of the experimental $^{89}$Y spectrum at $80K$ in O$%
_{6.6}$ to simulated ones and their corresponding polarization (inset).\
Dotted line indicates the experimental satellite shift. Polarizations $A$, $B
$ and $C$ have the same amplitude at $r=1$ but different curvatures. They
lead to the computed spectra $A$, $B$, and $C$. For this given amplitude,
only the $B$ exponential decay account for the experimental features. For a
different $r=1$ amplitude, the polarization $D$ leads to simulated spectrum $%
D$ which can also account for the experimental shape. On these simulated
spectra, the first satellite position does not coincide with the
experimental, as we use the value $^{89}A_{hf}\;$of the pure compound for
the hyperfine coupling between $^{89}$Y and $^{63}$Cu nn of the Li{\em .\ }

FIGURE 5 : Comparison of the experimental $^{17}$O spectrum (thick gray
line) at $80K$ in O$_{6.6}$ with $x_{N}=2\%$ of Li to the simulated ones
with the corresponding polarizations in insert, with same symbols. The short
range polarization is fixed to the $B$ case of Fig. 4. The shape of the
polarization for $r>\sqrt{5}$ is varied to fit the experimental spectrum. A
convolution of the computed spectra by a Lorentzian function corresponding
to the broadening of the pure sample has been performed. 

FIGURE 6 : Comparison of the experimental $^{17}$O spectrum at $80K$ in O$%
_{6.6}$ with $x_{N}=2\%$ of Li to simulated ones (b) with the corresponding
polarizations in (a), with the same symbols. For these different amplitudes $%
D$, $B$ and $E$, the best fit of the experimental spectrum is obtained with
an exponential curvature. Among these three cases, only $B$ can account both
for the shape and the broadening of the spectrum. 

FIGURE 7 : Polarization extension $\xi _{imp}$ (as defined in the text) as
function of temperature for underdoped O$_{6.6}$ (closed circles) and
slightly overdoped O$_{7}$ (open circles). The inset shows the polarization
envelope at $80K$ for O$_{6.6}$ and O$_{7}$. 

FIGURE 8 : Comparison of our result for the polarization extension $\xi
_{imp}$\ to the correlation length $\xi $ in pure compounds extracted from
INS\ experiments (\cite{BourgesBalatski}) and the MMP analysis (\cite
{BarzykinPines}). We plot also at O$_{7}$ the characteristic length $\xi _{n}
$ of the additional AF fluctuations observed in presence of Zn on the INS
spectra (\cite{Sidis}).


\bigskip

\begin{references}
\bibitem{Kampf}  \bigskip A.P. Kampf, Phys. Rep. {\bf 249}, 219 (1994)

\bibitem{BourgesBalatski}  A.V. Balatski and P.Bourges, Phys. Rev. Lett. 
{\bf 82}, 5337 (1999)

\bibitem{BarzykinPines}  A.J. Millis, H. Monien, and D. Pines, Phys. Rev. B 
{\bf 42}, 167 (1990); V. Barzykin and D. Pines, Phys. Rev. B {\bf 52}, 13585
(1995)

\bibitem{Mahajan}  A.V. Mahajan, H. Alloul, G. Collin and J.F. Marucco,
Phys. Rev. Lett. {\bf 72}, 3100 (1994)

\bibitem{JBobroff1997}  J. Bobroff, H. Alloul, Y. Yoshinari, A. Keren, P.
Mendels, N. Blanchard, G. Collin and J.-F. Marucco, Phys. Rev. Lett. {\bf 79}%
, 2117 (1997)

\bibitem{MarcHenriJ}  M.-H. Julien, T. Feh\'{e}r, M. Horvatic, C. Berthier,
O.N. Bakharev, P. S\'{e}gransan, G. Collin and J.-F. Marucco, Phys. Rev.
Lett. {\bf 84}, 3422 (2000)

\bibitem{Tedoldi}  F. Tedoldi, R. Santachiara and M. Horvatic, Phys. Rev.
Lett. {\bf 83}, 412 (1999); J.\ Das, A.\ Mahajan, J.\ Bobroff, H.\ Alloul,%
{\it \ }Phys. Rev. B {\bf 69}, 144404 (2004)

\bibitem{TakigawaFujiwara}  M. Takigawa, N. Motoyama, H. Eisaki, and S.
Uchida, Phys. Rev. B {\bf 55}, 14129 (1997); N. Fujiwara, H. Yasuoka, Y.
Fujishiro, M. Azuma, and M. Takano, Phys. Rev. Lett. {\bf 80}, 604 (1998)

\bibitem{JBPRLLi}  J. Bobroff, W.A. MacFarlane, H. Alloul, P. Mendels, N.
Blanchard, G. Collin and J.-F. Marucco, Phys. Rev. Lett. {\bf 83}, 4381
(1999)

\bibitem{MilaRice}  F. Mila and T.M. Rice, Physica C {\bf 157}, 561 (1989)

\bibitem{hyperfin}  We used for the hyperfine couplings the values :$%
^{17}A_{hf}^{c}=35.5kOe$, $^{89}A_{hf}^{c}=2.0kOe$, $^{63}A_{hf}^{c}=-88kOe$%
\ and $B=22kOe$ which are determined with a typical accuracy of 10\%\cite
{BarzykinPines}.\ This standard deviation mainly affects the value of the
hyperfine field of $^{7}$Li in the forthcoming analysis.

\bibitem{ITOH}  Y. Itoh, T. Machi, C. Kasai, S. Adachi, N. Watanabe, N.
Koshizuka and M. Murakami, Phys. Rev. B {\bf 67}, 64516 (2003)

\bibitem{PMendels}  P. Mendels, J. Bobroff, G. Collin, H. Alloul, M. Gabay,
J.-F. Marucco, N. Blanchard and B. Grenier, Europhys. Lett. {\bf 46}, 678
(1999)

\bibitem{BoyceSlichter}  J.B. Boyce and C.P. Slichter, Phys. Rev. Lett. {\bf %
32}, 61 (1974)

\bibitem{Sidis}  Y. Sidis, P. Bourges, B. Hennion, L. P. Regnault, R.
Villeneuve, G. Collin and J.-F. Marucco, Phys. Rev. B {\bf 53}, 6811 (1996)

\bibitem{PRLAndrew}  W. A. MacFarlane, J. Bobroff, H. Alloul, P. Mendels, N.
Blanchard, G. Collin and J.-F. Marucco, Phys. Rev. Lett. {\bf 85}, 1108
(2000)

\bibitem{footnoteTAU}  We used here the data for $\tau $ given in ref \cite
{PRLAndrew} at 100K, and recalling that $h/\tau \varpropto $ $^{7}A_{hf}$,
corrected by the ratio $1/2.5\;$of the hyperfine fields determined here and
estimated in ref \cite{PRLAndrew}

\bibitem{theseSidis}  Y. Sidis, P. Bourges, B. Keimer, unpublished data
(private communication)

\bibitem{WeakCorrelations}  N. Bulut, Physica {\bf 363C}, 260 (2001); Y.
Ohashi, J. Phys. Soc. Jpn. {\bf 70}, 2054 (2001)

\bibitem{StrongCorr}  A.M. Finkel'stein, V.E. Kataev, E.F. Kukovitskii and
G.B. Teitel'baum, Physica (Amsterdam) {\bf 168C}, 370 (1990); M. Gabay,
Physica {\bf 235-240C}, 1337 (1994); D. Poilblanc, D.J. Scalapino, and W.
Hanke, Phys. Rev. Lett. {\bf 72}, 884 (1994); N. Nagaosa and T.-K. Ng, Phys.
Rev. B {\bf 51}, 15588 (1995); R. Kilian, S. Krivenko, G. Khaliullin and P.
Fulde, Phys. Rev. B {\bf 59}, 14432 (1999)

\bibitem{MorrIncomT2}  J. Haase, D.K. Morr and C.P. Slichter, Phys. Rev. B 
{\bf 59}, 7191 (1999)

\bibitem{EffetplansT2}  A.J. Millis and H. Monien, Phys. Rev. B {\bf 54},
16172 (1996); A. Goto, W. G. Clark, P. Vonlanthen, K.B. Tanaka, T. Shimizu,
K. Hashi, P. V. P. S. S. Sastry and J. Schwartz, Phys. Rev. Lett. {\bf 89},
127002 (2002)

\bibitem{AeppliLSCO}  G. Aeppli, T.E. Mason, S.M. Hayden, H.A. Mook and J.
Kulda, Science {\bf 278}, 1432 (1997)

\bibitem{Markiewicz}  R.S. Markiewicz, cond-mat/0312595

\bibitem{Mook}  H.A. Mook, P. Dai, S.M. Hayden, G. Aeppli, T.G. Perring and
F. Dogan, Nature {\bf 395}, 580 (1998)
\end{references}
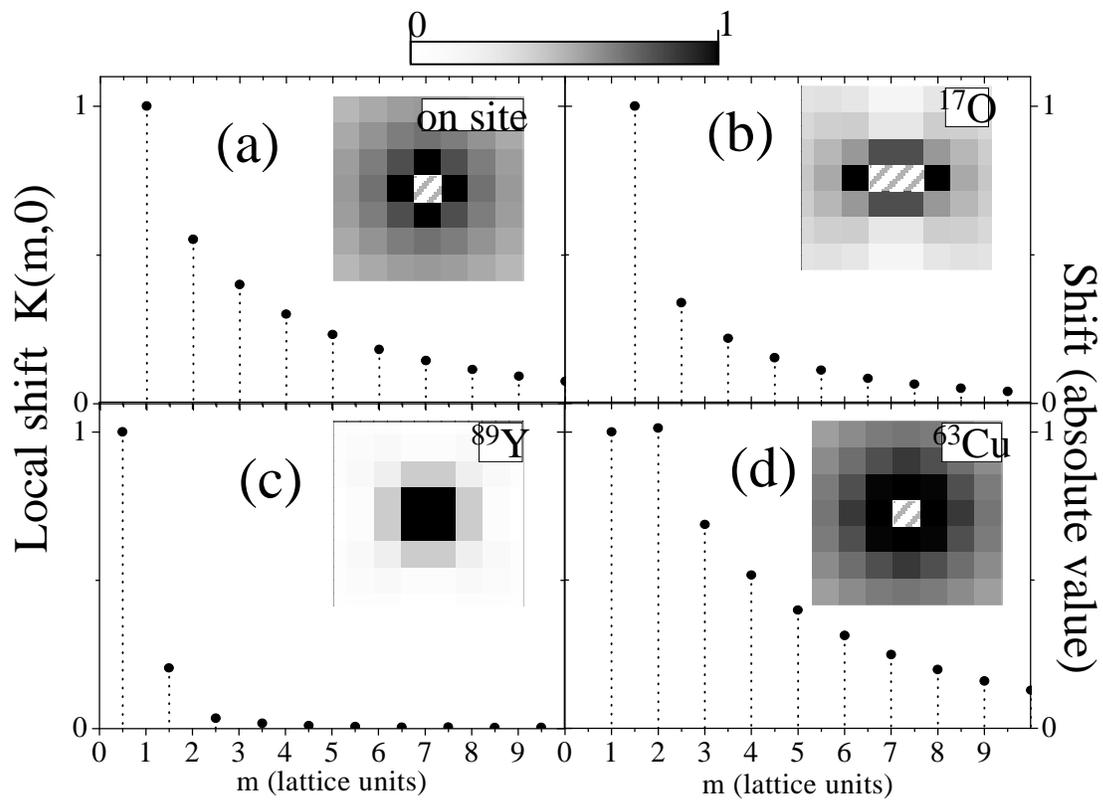
\end{document}